\documentstyle[amssymb,psfig,deluxetable]{mn2e}

\title[No compelling evidence of significant early star cluster
  disruption in the LMC]{No compelling evidence of significant early
  star cluster disruption in the Large Magellanic Cloud}

\author[Richard de Grijs, Simon P. Goodwin and Peter Anders]{Richard
  de Grijs,$^{1,2}$\thanks{E-mail: grijs@pku.edu.cn} Simon
  P. Goodwin$^3$ and Peter Anders$^{1,4}$\\
$^1$ Kavli Institute for Astronomy and Astrophysics, Peking
  University, Yi He Yuan Lu 5, Hai Dian District, Beijing 100871,
  China\\
$^2$ Department of Astronomy, Peking University, Yi He Yuan Lu 5, Hai
  Dian District, Beijing 100871, China\\
$^3$ Department of Physics \& Astronomy, The University of Sheffield,
  Hicks Building, Hounsfield Road, Sheffield S3 7RH\\
$^4$ Key Laboratory for Optical Astronomy, National Astronomical
  Observatories, Chinese Academy of Sciences, 20A Datun
  Road,\\ Chaoyang District, Beijing 100012, China}

\date{Received date; accepted date}
\pubyear{2013}

\begin{document}
\maketitle

\begin{abstract}
Whether or not the rich star cluster population in the Large
Magellanic Cloud (LMC) is affected by significant disruption during
the first few $\times 10^8$ yr of its evolution is an open question
and the subject of significant current debate. Here, we revisit the
problem, adopting a homogeneous data set of broad-band imaging
observations. We base our analysis mainly on two sets of
self-consistently determined LMC cluster ages and masses, one using
standard modelling and one which takes into account the effects of
stochasticity in the clusters' stellar mass functions. On their own,
the results based on any of the three complementary analysis
approaches applied here are merely indicative of the physical
conditions governing the cluster population. However, the combination
of our results from all three different diagnostics leaves little room
for any conclusion other than that the optically selected LMC star
cluster population exhibits no compelling evidence of significant
disruption -- for clusters with masses, $M_{\rm cl}$, of $\log( M_{\rm
  cl}/{\rm M}_\odot) \gtrsim 3.0$--3.5 -- between the age ranges of
[3--10] Myr and [30--100] Myr, either `infant mortality' or
otherwise. In fact, there is no evidence of any destruction beyond
that expected from simple models just including stellar dynamics and
stellar evolution for ages up to 1 Gyr. It seems, therefore, that the
difference in environmental conditions in the Magellanic Clouds on the
one hand and significantly more massive galaxies on the other may be
the key to understanding the apparent variations in cluster disruption
behaviour at early times.
\end{abstract}


\section{Introduction}
\label{intro.sec}

Star clusters are the most highly visible stellar population
components in galaxies beyond the Local Group. Their integrated
properties are generally used to trace, e.g., their host galaxy's star
(cluster) formation history, the impact and time-scales of the most
recent (major) mergers or close encounters with any neighbouring
galaxies, and the extent to which environmental conditions drive the
evolution of star cluster systems in their own right.

The galaxies in the Local Group represent unique benchmarks which can
be used to verify analyses based on integrated cluster properties
using resolved stellar photometry (e.g., de Grijs \& Anders 2006;
Colucci \& Bernstein 2012; Baumgardt et al. 2013; Cezario et al. 2013;
de Meulenaer et al. 2013). As such, the star cluster systems in the
Small and Large Magellanic Clouds (SMC, LMC) can provide unique
insights into the properties of their resolved star cluster
populations. Prompted by recent claims (Chandar, Fall \& Whitmore
2010a; Chandar, Whitmore \& Fall 2010b) and counterclaims (e.g.,
Baumgardt et al. 2013) that the disruption rate of star clusters in
the LMC may be significant from early ages (a few Myr) up to an age of
$\sim 1$ Gyr, we decided to revisit this issue based on a number of
complementary approaches.

Chandar et al. (2010a,b) determined the cluster population's age and
mass distributions based on fits to the broad-band spectral-energy
distributions (SEDs) from Hunter et al.'s (2003) comprehensive
database of integrated LMC cluster photometry. They used simple
stellar population (SSP) models characterized by fully sampled stellar
mass functions (MFs) for their parameter derivation (cf. Section
2). Two other studies used exactly the same photometric database to
independently determine the clusters' ages and masses. Specifically,
de Grijs \& Anders (2006) adopted fully sampled SSP models to
determine the LMC cluster population's properties, while Popescu et
al. (2012) took into account the effects of stochastically sampled
stellar MFs, which become particularly noticeable for cluster masses
below a few $\times 10^4$ M$_\odot$ (cf. Section 3). Given that these
studies all used the same basic cluster photometry, it is instructive
to first compare Chandar et al.'s (2010a,b) results with those of de
Grijs \& Anders (2006), since both teams based their parameter
determinations on the same underlying physical assumptions (barring
small differences between the actual SSP models used, which we discuss
below where relevant). We will then proceed by properly taking into
account the effects of stochastic sampling of the clusters' stellar
MFs, which is arguably a physically sounder assumption for lower-mass
clusters.

This debate goes beyond the mere niche of the question as to how star
cluster populations evolve. Most importantly, it touches upon the
process in which disrupting star clusters populate their host galaxy's
galactic field. At present, two competing theories hold sway in this
area. One supports the idea that early star cluster disruption is
independent of cluster mass and does not depend on the clusters'
environment either (e.g., Chandar et al. 2010a,b; Fall \& Chandar
2012), which must be contrasted with the view that environmental
differences lead to different cluster disruption signatures, which may
also exhibit a dependence on cluster mass (e.g., de Grijs \& Goodwin
2008, 2009, and references therein; see also Lamers 2009). In this
paper we will show that, at least for the LMC and for the data set in
common among all competing studies (Hunter et al. 2003; de Grijs \&
Anders 2006; Chandar et al. 2010a,b; Popescu et al. 2012; Baumgardt et
al. 2013), the overwhelming evidence rules out -- at high statistical
significance -- substantial cluster disruption at early times ($t
\lesssim 10^8$ yr). In Section \ref{bigpicture.sec}, we will place
these results in a more general context.

\section{Cluster data}
\label{data.sec}

In de Grijs \& Anders (2006) we compared the physical parameters of
the LMC's star cluster population obtained from resolved photometry
and spectroscopy on the one hand and integrated SEDs on the
other. This was necessarily restricted to age comparisons of the more
massive LMC sample clusters, as constrained by the availability of
prior age determinations in the literature at that time. Using our
{\sc AnalySED} tool for star cluster analysis based on broad-band SEDs
assuming fully populated stellar MFs (Anders et al. 2004b), we
re-analysed the current most comprehensive database of integrated LMC
cluster photometry (Hunter et al. 2003).

Prior to this, we had already concluded (de Grijs et al. 2005) that
application of the {\sc AnalySED} approach based on standard modelling
employing the {\sc galev} SSP models (Kotulla et al. 2009; and
references therein, as well as subsequent, unpublished updates) showed
that the {\it relative} masses within a given cluster system can be
determined to very high accuracy (provided that the clusters' stellar
MFs are well-populated), depending on the specific combination of
passbands used (Anders et al. 2004b). Under the conditions explored in
de Grijs et al. (2005), we found that the {\it absolute} accuracy with
which the cluster mass distribution can be reproduced using different
model approaches (including different SSP models, filter combinations,
and input physics) is $\sigma_M = \Delta \langle \log( M_{\rm cl} /
{\rm M}_\odot) \rangle \le 0.14$, compared with $\sigma_t = \Delta
\langle \log( t \mbox{ yr}^{-1} ) \rangle \le 0.35$ for the age
distribution: ``[t]his implies that mass determinations are mostly
insensitive to the approach adopted'' (de Grijs et al. 2005), because
the mass-to-light ratio of a given SSP depends only weakly on the
population's age, at least within reasonably narrow age ranges. In any
cluster analysis used to derive ages and masses, the age uncertainties
are, by far, the most significant.

Driven by a number of controversies that had appeared in the
literature (Chandar, Fall \& Whitmore 2006; Gieles, Lamers \&
Portegies Zwart 2007), we proceeded to apply our analysis approach to
the SMC's star cluster system (de Grijs \& Goodwin 2008), based on
Hunter et al.'s (2003) broad-band magnitudes. We concluded that the
optically selected SMC star cluster population has undergone at most
$\sim 30$ per cent disruption between the age ranges of approximately
[3--10] Myr and [40--160] Myr, a process often referred to as `infant
mortality'.\footnote{When we refer to `infant' mortality in this
  paper, this relates to the mass-independent disruption of a fraction
  of the total cluster sample owing to rapid gas expulsion during the
  first $\sim 10^7$ yr (with an upper limit of 2--$4 \times 10^7$ yr)
  of the population's lifetime.} We ruled out an alleged (Chandar et
al. 2006) 90 per cent cluster disruption rate per decade of $\log( t
\mbox{ yr}^{-1})$ for $t \le 10^9$ yr.

In the mean time, Chandar et al. (2010a,b) have used the same Hunter
et al. (2003) photometric database, combined with their independently
determined yet unpublished age and mass estimates for 854 LMC
clusters, to raise a new controversy. They suggest that a scenario in
which clusters undergo gradual, mass-independent disruption up to $t
\sim 1$ Gyr provides the best match to the data. However, Baumgardt et
al. (2013) recently concluded that significant cluster disruption
appears to set in only after an age of $\sim 200$ Myr (see also de
Grijs \& Goodwin 2009). This latter conclusion is consistent with the
results of Parmentier \& de Grijs (2008). Neither of these latter
authors focussed on the evolution of the youngest clusters, however,
nor did Baumgardt et al. (2013) explore the apparent discrepancies
with Chandar et al. (2010a,b) in detail. Addressing these two aspects
is what we set out to do here.

In addition, upon close inspection of the LMC cluster database (which
was kindly provided by D. Hunter), it turns out that it includes a
significant number of duplicate clusters. These duplicates were not in
all cases identified by Hunter et al. (2003), but they only become
apparent based on a detailed comparison of the spatial distribution of
the clusters. We hence proceeded to clean the Hunter et al. (2003)
database, resulting in a sample of 748 unique clusters (see also
Popescu, Hanson \& Elmegreen 2012; Baumgardt et al. 2013).

\begin{figure}
\psfig{figure=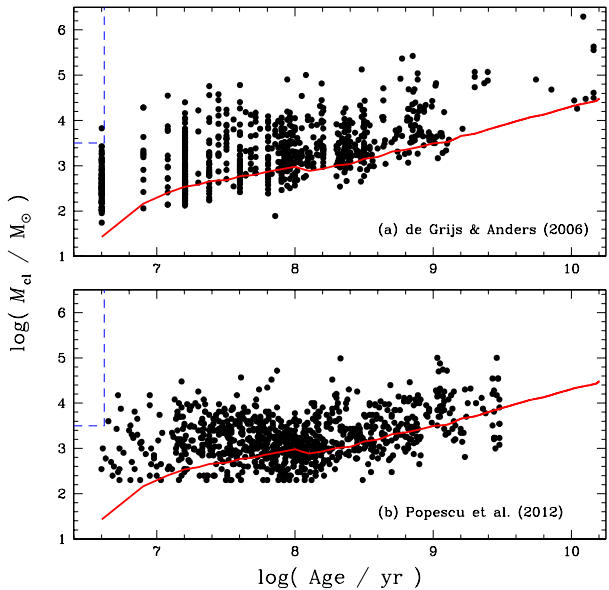,width=\columnwidth}
\caption{\label{amd.fig}(a) Age--mass distribution (de Grijs \& Anders
  2006) of the 748 optically selected LMC clusters based on our
  updated database. (b) As panel (a), but for the parameters
  determined by Popescu et al. (2012; their tables 1 and 2) for 920
  LMC clusters. For reasons of presentational clarity we have omitted
  the relevant error bars on the data points in both panels, although
  they have been taken into account properly in our analysis (see
  text). The solid (red) lines indicate the approximate 50 per cent
  completeness limits, $M_V \simeq -4.3$ mag, based on the {\sc galev}
  SSP models. The blue dashed boxes indicate a section of parameter
  space which will be discussed in Section \ref{discrepancies.sec}.}
\end{figure}

\begin{table*}
\caption[ ]{\label{tab1}LMC cluster positions and derived parameters.}
\begin{center}
{\scriptsize
\begin{tabular}{ccccccccccl}
\hline
\hline
RA (J2000) & Dec (J2000) & $M_V$ & & $\log(t \mbox{ yr}^{-1})$
& & & & $\log(M_{\rm cl}/{\rm M}_\odot)$ & & Cluster name(s) \\
\cline{4-6} \cline{8-10} 
{(hh mm ss.ss)} & {(dd mm ss.ss)} & {(mag)} & {(min)} & {(best)} &
{(max)} & & {(min)} & {(best)} & {(max)}\\
\hline
04 44 47.00 & $-$69 38 31.83 & $-$4.120 & 8.134 & 8.318 & 8.358 && 3.009 & 3.161 & 3.190 & LW46, KMHK63 \\ 
04 44 59.63 & $-$70 18 05.40 & $-$3.807 & 7.944 & 8.000 & 8.301 && 2.730 & 2.764 & 2.998 & BSDL14 \\ 
04 45 05.91 & $-$68 47 43.18 & $-$1.641 & 6.903 & 7.857 & 8.301 && 1.029 & 1.892 & 2.220 & SL27, KMHK64 \\ 
04 45 08.87 & $-$69 48 11.20 & $-$2.624 & 7.964 & 8.326 & 8.422 && 2.267 & 2.542 & 2.611 & LW49, KMHK67 \\ 
04 45 41.00 & $-$70 59 23.00 & $-$3.789 & 8.064 & 8.334 & 8.408 && 2.791 & 3.000 & 3.053 & SL31, LW53, KMHK75 \\ 
$\cdots$ & $\cdots$ & $\cdots$ & $\cdots$ & $\cdots$ & $\cdots$ && $\cdots$ & $\cdots$ & $\cdots$ \\ 
\hline 
\end{tabular}
}
\end{center}
\flushleft
Notes:\\ The $1\sigma$ uncertainties in the age and mass estimates are
represented by the differences between the `best' values and the
minimum/ maximum allowable solutions from the analysis of de Grijs \&
Anders (2006), who adopted $Z = 0.4 \; {\rm Z}_\odot$ and $E(B-V)=0.1$
mag. Table \ref{tab1} is published in its entirety in the electronic
edition of the paper. A portion is shown here for guidance regarding
its form and content.
\end{table*}

Figure \ref{amd.fig}a shows the LMC cluster distribution in the
diagnostic age--mass diagram based on our cleaned database; the
relevant cluster parameters are included in Table \ref{tab1}. The
original integrated cluster photometry is available from Popescu et
al. (2012; their tables 1 and 2). At first glance, two narrow features
in the age distribution are apparent. These so-called `chimneys' at
$\log( t \mbox{ yr}^{-1}) = 6.6$ and $\sim 7.2$ are associated with,
respectively, the minimum age included in our SSP models (any clusters
characterized by younger SEDs are returned to the youngest age by our
fitting routines) and the onset of red supergiants in realistic
stellar populations. The latter chimney is an artefact caused by a
local minimum in parameter space. We also note that the observational
completeness limit (indicated by the solid red line, which represents
the $\sim 50$ per cent completeness level, at $M_V \simeq -4.3$ mag;
for a discussion, see de Grijs \& Anders 2006) is a function of age,
so that -- depending on the age range of interest -- one needs to vary
the minimum mass to compare and assess the MFs of different cluster
subsamples.

Finally, we explored whether any other existing databases of LMC
cluster parameters could be exploited to support the analysis
presented in this paper. We specifically focussed on the catalogue of
Glatt, Grebel \& Koch (2010), who compiled data of 1193 populous LMC
clusters with ages of up to 1 Gyr based on the most up-to-date and
comprehensive LMC object catalogue of Bica et al. (2008). Glatt et
al. (2010) used the optical broad-band photometry from the Magellanic
Clouds Photometric Survey (MCPS; Zaritsky et al. 2004) to construct
colour--magnitude diagrams (CMDs) and subsequently determined ages for
their entire sample based on isochrone fits. Unfortunately, the lower
age boundary pertaining to the Glatt et al. (2010) sample is poorly
defined. They only performed isochrone fitting of CMDs associated with
objects identified as genuine clusters (flagged `C') by Bica et
al. (1996). This selection resulted in poorly understood systematic
effects, however: (i) Bica et al.'s (1996) classification is,
essentially, based on visual examination and hence affected by
subjectivity, and (ii) very young objects are usually classified as
`associations' or `nebulae', which leads to a subjective, {\it
  variable} lower age limit of $\sim 10$ Myr to the Glatt et
al. (2010) sample. These considerations render the applicability of
the latter catalogue rather limited in the context of our assessment
of the reality of early star cluster disruption in the
LMC. Nevertheless, this database can and will be used to provide
circumstantial support to our results in Section \ref{disruption.sec}.

\section{Stochasticity in the clusters' stellar mass functions}

In analyses of integrated star cluster photometry, one must be careful
to assess the effects of stochastic sampling of the stellar initial MF
(IMF). Particularly for cluster masses $M_{\rm cl} \lesssim \mbox{ a
  few} \times 10^4 {\rm M}_\odot$, broad-band SEDs may yield
significantly different ages and -- to a lesser extent -- masses than
the true cluster parameters (e.g., Cervi\~no, Luridiana \& Castander
2000; Cervi\~no et al. 2002; Cervi\~no \& Luridiana 2004, 2006;
Barker, de Grijs \& Cervi\~no 2008; Ma\'{\i}z Apell\'aniz 2009;
Popescu \& Hanson 2010; Fouesneau \& Lan\c{c}on 2010; Silva-Villa \&
Larsen 2010, 2011; Fouesneau et al. 2012; Popescu et al. 2012; Anders
et al. 2013). Since our cluster mass estimates go down to a few
$\times 10^3 {\rm M}_\odot$, one should expect that our results would
also be affected by stochasticity, although we point out that in de
Grijs \& Anders (2006) we found excellent agreement between our age
estimates based on broad-band SED analysis and those from resolved
photometry or spectroscopy, provided that the effects of the
age--extinction and age--metallicity degeneracies are duly taken into
account.

Recently, Popescu et al. (2012) re-analysed the LMC cluster photometry
of Hunter et al. (2003) using their novel {\sc massclean}{\it age}
approach, which allows one to take into account the effects of
stochastic sampling of the stellar MF and, hence, determine the
uncertainties associated with adoption of such MFs. The age--mass
diagram based on their modelling is shown in
Fig. \ref{amd.fig}b. Although both panels of Fig. \ref{amd.fig} show
appreciable differences in the details, the overall distributions
appear fairly similar in terms of their coverage of the relevant
parameter space, particular once one considers the clusters well above
the 50 per cent completeness limit (e.g., both catalogues are roughly
equally split into clusters younger and older than 100 Myr). Most
importantly in the context of the present work, the Popescu et
al. (2012) results yield significantly lower masses for a fraction of
the LMC clusters (i.e., those located below the generic 50 per cent
completeness limit), which is an expected effect of fitting integrated
magnitudes affected by stochastically sampled stellar MFs with fully
sampled SEDs (e.g., Silva-Villa \& Larsen 2010, 2011; Anders et
al. 2013). We also note that Popescu et al.'s (2012) cluster ages
extend up to $\log( t \mbox{ yr}^{-1}) = 9.5$, while the {\sc galev}
models used to construct Fig. \ref{amd.fig}a include older ages.
Close inspection of both sets of results shows that of the 10 clusters
rendered older than $\log( t \mbox{ yr}^{-1}) = 9.5$ by our {\sc
  galev}-based approach, six and four were returned as, respectively,
$\log( t \mbox{ yr}^{-1}) \sim 9$ and $\log( t \mbox{ yr}^{-1}) < 8$
by the {\sc massclean}{\it age} approach.

Baumgardt et al. (2013; their fig. 2) compared the age determinations
of de Grijs \& Anders (2006) with those of Popescu et al. (2012) and
found a systematic deviation from the one-to-one locus. They suggested
that this tilt in the distribution is most likely caused by the
effects of stochasticity. Here, we make an effort at quantifying the
impact of stochastic effects, since this will be important for the
discussion in the remainder of the paper. In Table \ref{tab2} we
compare the slope (including the statistical uncertainty in the fit)
in the $\log( t \mbox{ yr}^{-1})$ [de Grijs \& Anders 2006] versus
$\log( t \mbox{ yr}^{-1})$ [Popescu et al. 2012] diagram for (i)
different low-mass limits and (ii) using both the de Grijs \& Anders
(2006) and the Popescu et al. (2012) cluster mass determinations as
our basis. It appears that for cluster masses $\log(M_{\rm cl} / {\rm
  M}_\odot) \gtrsim 3.0$--3.5 (where the value of the lower limit
depends on the database used for the mass determination), the age
comparison is statistically consistent with a one-to-one
distribution. Among the clusters younger than $10^9$ yr (the age range
of interest in this paper) in the de Grijs \& Anders (2006) sample,
142 of 550 (25.8 per cent) are less massive than $\log(M_{\rm cl}/{\rm
  M}_\odot) = 3.0$ yet brighter than the canonical selection limit at
$M_V = -4.3$ mag. A similar fraction, 21.6 per cent (119 of 552
clusters), meet the same selection criteria in the Popescu et
al. (2012) sample.

\begin{table*}
\caption[ ]{\label{tab2}Quantitative, statistical comparison of the
  derived LMC cluster ages for different sample selections.}
\begin{center}
\begin{tabular}{lcccc}
\hline
\hline
\multicolumn{1}{c}{Sample} & Reference$^a$ & Slope$^b$ & Reference$^a$ & Slope$^b$ \\
\hline
All clusters                               & N/A                       & $0.82 \pm 0.02$ & N/A                   & $0.82 \pm 0.02$ \\
$\log(M_{\rm cl} / {\rm M}_\odot) \ge 3.0$ & de Grijs \& Anders (2006) & $0.96 \pm 0.04$ & Popescu et al. (2012) & $0.90 \pm 0.03$ \\
$\log(M_{\rm cl} / {\rm M}_\odot) \ge 3.5$ & de Grijs \& Anders (2006) & $1.03 \pm 0.06$ & Popescu et al. (2012) & $1.03 \pm 0.03$ \\
$\log(M_{\rm cl} / {\rm M}_\odot) \ge 4.0$ & de Grijs \& Anders (2006) & $1.00 \pm 0.11$ & Popescu et al. (2012) & $1.07 \pm 0.04$ \\
\hline
\end{tabular}
\end{center}
\flushleft 
Notes:\\ $^a$ Database used to determine the lower mass limit; $^b$
Horizontal axis: age determinations from de Grijs \& Anders (2006);
vertical axis: age determinations from Popescu et al. (2012).
\end{table*}

\begin{figure}
\begin{center}
\psfig{figure=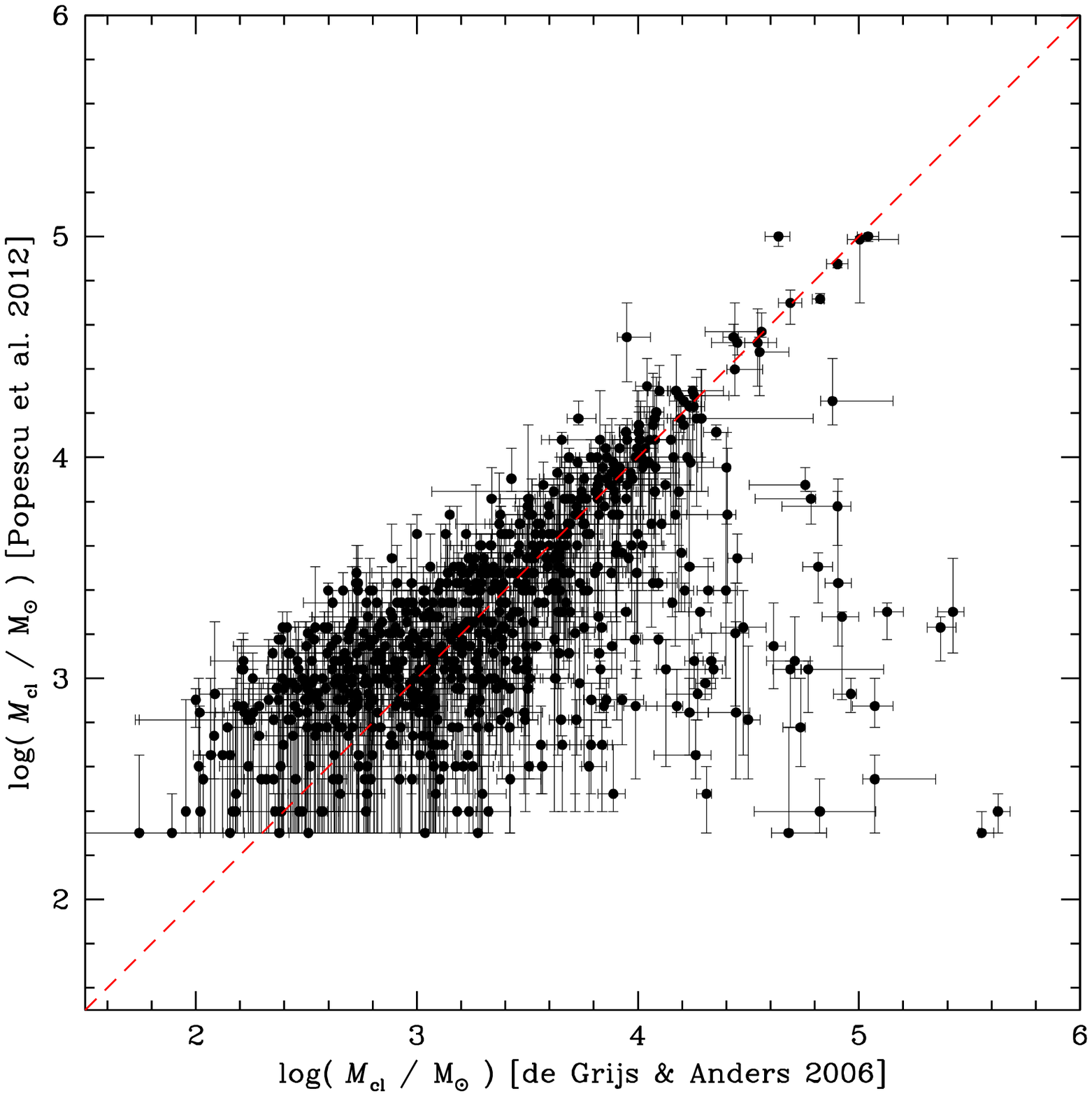,width=\columnwidth}
\end{center}
\begin{center}
\psfig{figure=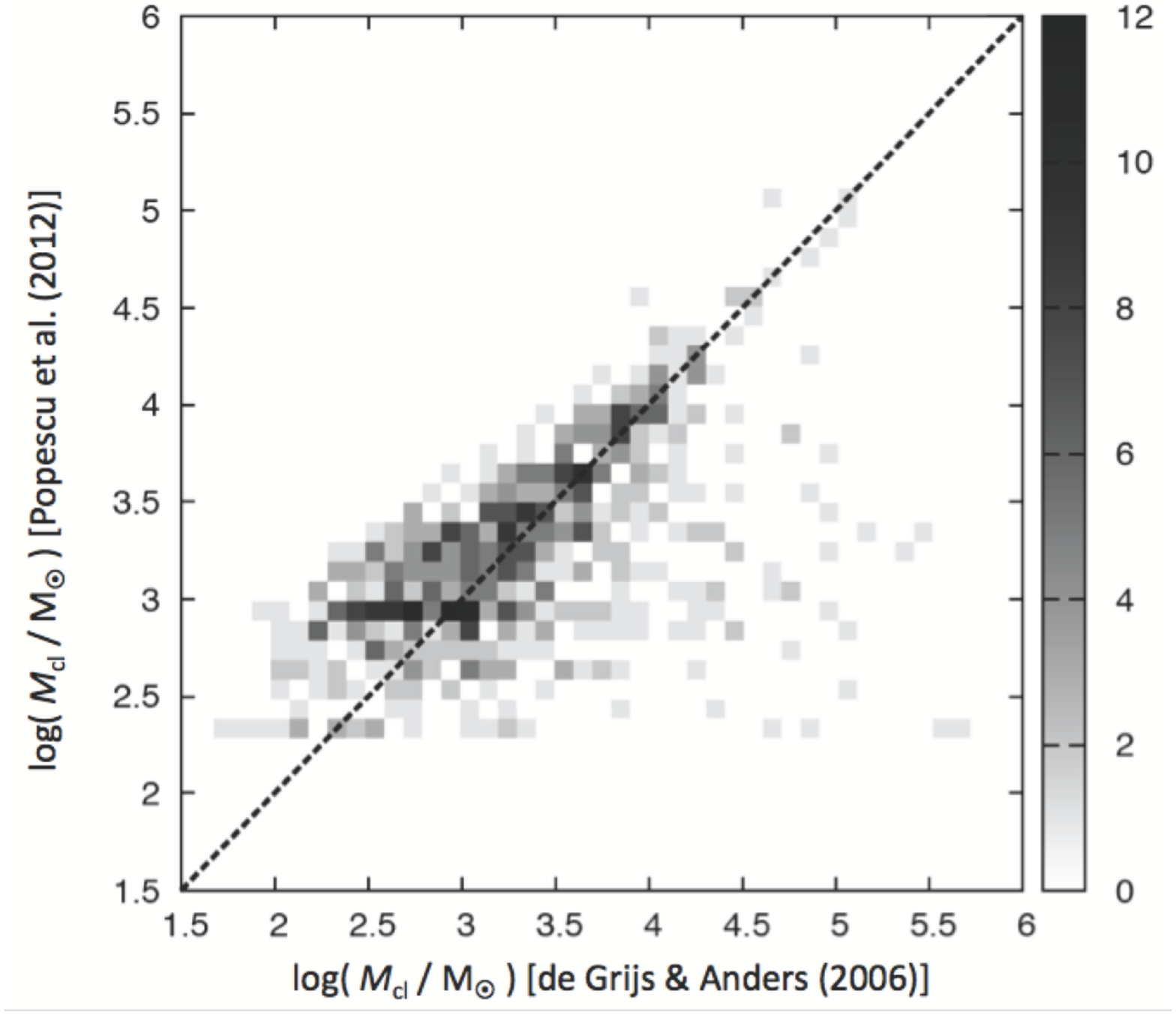,width=0.9\columnwidth}
\end{center}
\caption{\label{masscf.fig}Comparison of cluster masses determined by
  de Grijs \& Anders (2006) and Popescu et al. (2012). Top: Direct
  comparison, including uncertainties. Bottom: Representation as a
  density distribution.}
\end{figure}

This conclusion is also consistent with the statistical differences
between the cluster mass determinations. Fig. \ref{masscf.fig} shows
the extent to which the de Grijs \& Anders (2006) and Popescu et
al. (2012) masses are comparable. The top panel shows the individual
mass measurements and their associated uncertainties. The bottom panel
is based on the same data set, but here we use a density distribution
to highlight the locus of the majority of our sample clusters. It is
clear that, for clusters with $\log( M_{\rm cl} / {\rm M}_\odot )
\lesssim 3.5$, the Popescu et al. (2012) masses are systematically
higher than their counterparts from de Grijs \& Anders (2006). For
higher-mass clusters, the similarity between both studies is, in fact,
quite close. The masses and ages in both data sets are statistically
similarly distributed for clusters with $\log( M_{\rm cl} / {\rm
  M}_\odot ) \ge 3.5$, within the associated uncertainties. For
instance, for $\log( M_{\rm cl} / {\rm M}_\odot ) \ge 3.0$ (3.5), the
slope in Fig. \ref{masscf.fig} is $1.04 \pm 0.05$ $(1.00 \pm 0.08)$.

As a result of these considerations, we are confident that the effects
of stochasticity in the clusters' stellar MFs, while clearly present,
do not significantly impede our analysis. In the remainder of this
paper and where relevant, we will split up our sample of LMC clusters
into different mass-limited subsamples, to explore specifically
whether stochastic sampling effects could have a significant impact on
our conclusions. Finally, and perhaps most importantly, we also note
that the comparison studies using this same database (in particular
Chandar et al. 2010a,b) are similarly affected by these effects.  The
effects of taking into account stochastic sampling become clear when
we consider the numbers of young, $\le 10^9$ yr-old clusters between
$\log(M_{\rm cl}/{\rm M}_\odot) = 3.0$ and 3.5 in both of our
catalogues. We find a total of 179 clusters (32.5 per cent) of
clusters in this selection box in the de Grijs \& Anders (2006)
database, compared with 262 objects (47.6 per cent) in the Popescu et
al. (2012) tables.

\section{Cluster mass functions}
\label{cmf.sec}

In de Grijs \& Goodwin (2008) we explored the potential effects of
star cluster infant mortality in the SMC by analysing the cluster MFs
as a function of age. Particularly for the youngest ages, cluster MFs
are well described by power-law distributions of the form $N_{\rm cl}
\propto M_{\rm cl}^{-\alpha}$, where $N_{\rm cl}$ is the number of
clusters of mass $M_{\rm cl}$, while the power-law slope $\alpha$ is
usually close to 2 (e.g., de Grijs et al. 2003; Portegies Zwart,
McMillan \& Gieles 2010; Fall \& Chandar 2012). Here we apply the same
analysis techniques to our LMC sample. One significant advantage of
using the LMC cluster sample compared with the SMC cluster population
is its approximately threefold larger number of clusters, resulting in
comparatively smaller statistical uncertainties.

\begin{figure*}
\begin{tabular}{cc}
\psfig{figure=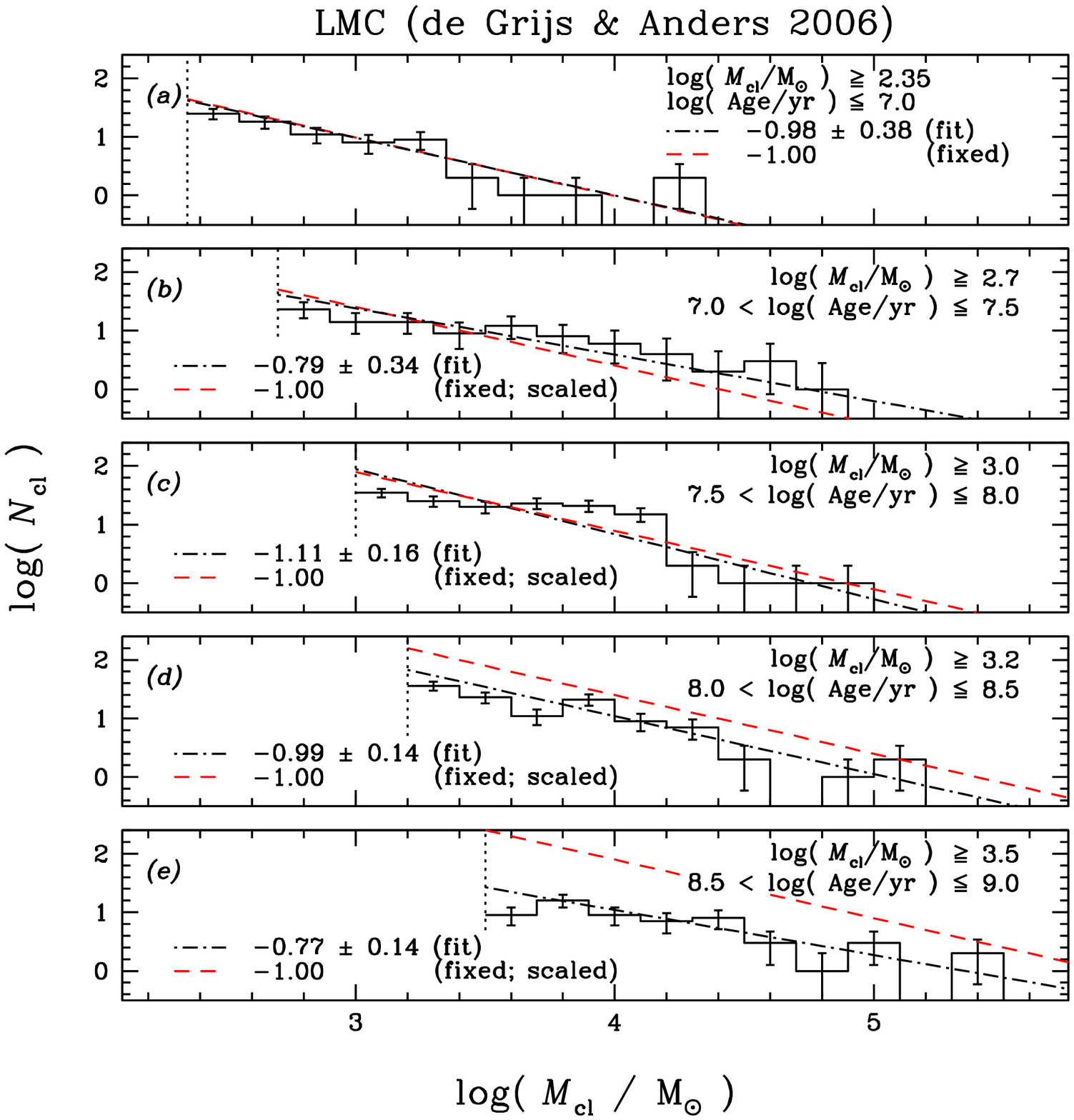,width=\columnwidth} &
\psfig{figure=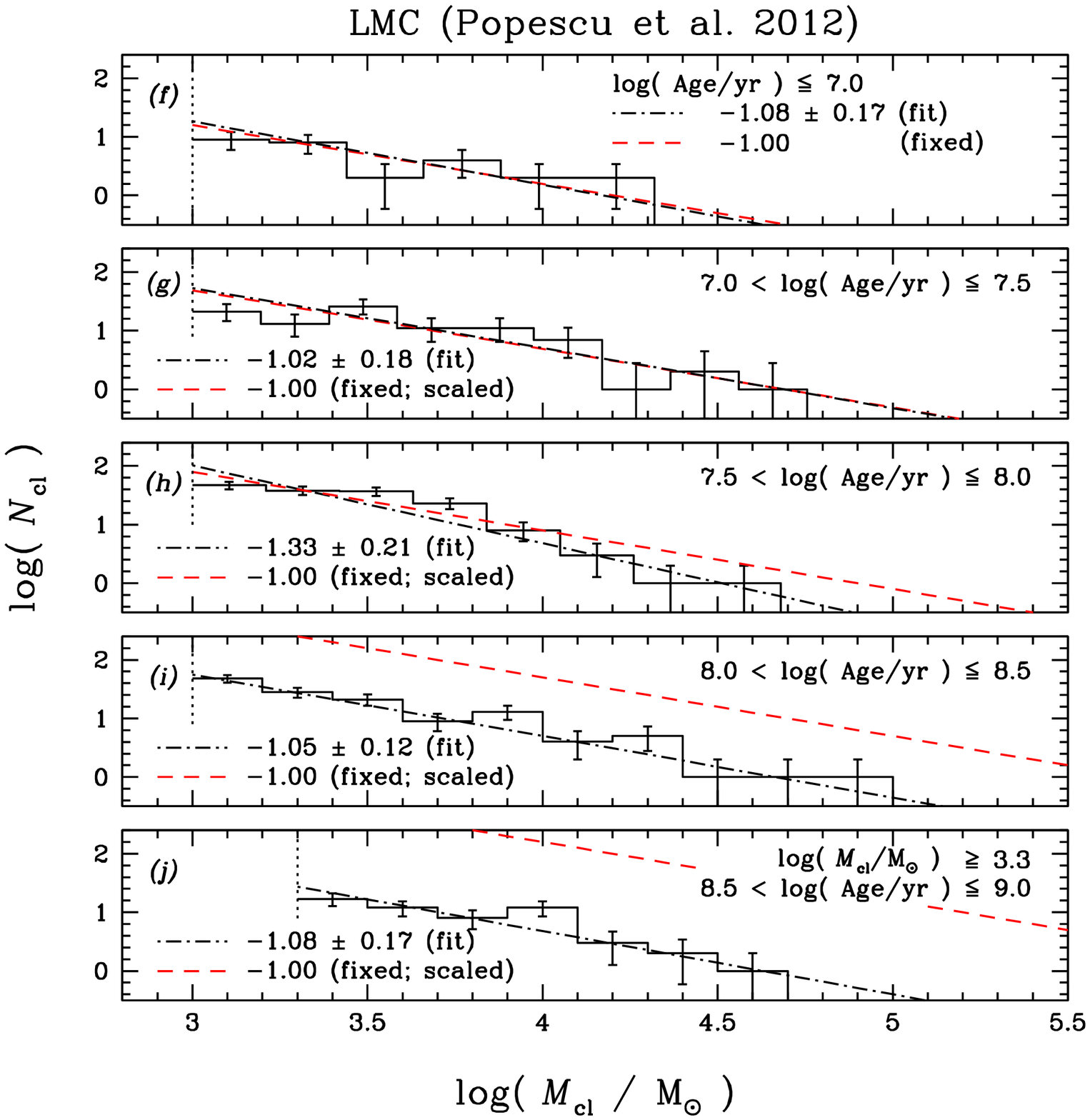,width=\columnwidth} \\
\end{tabular}
\caption{\label{lmcclf.fig}Cluster MFs for statistically complete LMC
  cluster subsamples. Age and mass ranges are indicated in most panel
  legends; for panels (f)--(i) we have adopted a minimum cluster mass
  of $10^3$ M$_\odot$. The vertical dotted lines indicate the low-mass
  limits adopted for the power-law distributions; for panels (a)--(e)
  and (i)--(j), these represent the approximate 50 per cent
  completeness limits. Error bars represent Poissonian errors, while
  the (red) dashed lines represent cluster MFs of `canonical' slope
  $\alpha = 2$, shifted vertically as described in the text. Except
  for the dashed line in the top panels, these canonical MFs are {\it
    not} fit results. The (black) dash-dotted lines represent the
  best-fitting cluster MFs for $3.0 \le \log(M_{\rm cl}/{\rm M}_\odot)
  \le 5.0$; for panels (d), (e) and (j) we used the selection limit as
  lower fitting boundary.}
\end{figure*}

Figure \ref{lmcclf.fig} shows the cluster MFs for five statistically
complete LMC cluster subsamples, based on both the de Grijs \& Anders
(2006) and Popescu et al. (2012) age and mass determinations (left-
and right-hand columns, respectively). We have included the best
power-law fits as black dash-dotted lines. Note that in the
representation where we show $\log(N_{\rm cl})$ as a function of
$\log(M_{\rm cl})$, the canonical power-law index of $-2$ translates
into a slope of $-1$. It is clear that for $\log(t \mbox{ yr}^{-1})
\lesssim 8.0$--8.5 (depending on the parameter set used for the
analysis) the MFs are well described by such a canonical power-law
function. The red dashed lines represent these power laws with a slope
of $-1$ in the parameter space defined by Fig. \ref{lmcclf.fig}. In
panels (b)--(e) and (g)--(j) we show the canonical MFs, scaled from
the best-fitting loci in Figs \ref{lmcclf.fig}a and \ref{lmcclf.fig}f,
respectively, by the difference in age range between the panels. In de
Grijs \& Goodwin (2008), we explained that the main uncertainties
introduced by adopting this method are owing to fluctuations caused by
small-number statistics in the youngest age range and the exact length
of the youngest age range, for which we adopted a minimum age for
optically visible clusters of 3 Myr. The youngest age limit is set by
the time it takes a cluster to emerge from its natal gas and dust
cloud and become optically visible (cf. de Grijs \& Goodwin 2008).

The scaled canonical cluster MFs provide remarkably good matches to
the MFs in, respectively, panels (b)--(c) and (g)--(h), given the
simplifying underlying (null) hypothesis of constant cluster
formation. The small apparent difference between the canonical and
best-fitting slopes in panel (b) -- although they are still comparable
within the formal statistical uncertainties -- is likely owing to the
appearance of red supergiants in this age range, combined with the
possible effects of stochasticity. (The presence of red supergiants in
stochastically sampled clusters will cause SED fits based on fully
sampled IMFs to return cluster masses that are biased towards higher
values.) Stochastic sampling effects are also the likely cause for the
$\sim 1$--$2 \sigma$ slope discrepancy seen in panel (h).  Based on
this analysis alone, it appears that the effects of significant
cluster disruption become apparent only beyond $\log(t \mbox{
  yr}^{-1}) \sim 8$. Although we do not claim that this result on its
own validates the assumption of constant cluster formation, nor the
absence of rapid cluster disruption in the LMC, it contributes to the
overall, self-consistent picture of early cluster evolution which we
are painting in this paper.

Under the assumption that the cluster formation rate has remained
roughly constant (within 10 per cent for $t \le 10^9$ yr;
cf. Maschberger \& Kroupa 2011; see also Section
\ref{bigpicture.sec}), we conclude on the basis of
Fig. \ref{lmcclf.fig} that there is no compelling evidence of
significant mortality, either infant mortality or disruption up to
$\sim 100$ Myr and within the Poissonian uncertainties. In the next
section, we will attempt to quantify the maximum disruption rate
allowed by the data and the corresponding uncertainties.

\section{Disruption or evolution?}
\label{disruption.sec}

To underscore the key result from the previous section, in
Fig. \ref{dndt.fig} we plot the LMC cluster age distribution expressed
in number of clusters per Myr. We show both the full,
magnitude-limited LMC cluster sample and three mass-limited
subsamples. In addition, we have included two arrows to highlight the
main differences between the theoretical expectations for no cluster
infant mortality and a 90 per cent disruption rate per decade in
age. Both predictions also include the usual effects of stellar
evolution and fading, i.e., they follow standard SSP evolution as
implemented in the {\sc galev} models. Just as for the SMC cluster
system, if we force it to pass through the data point associated with
the youngest age range, the blue, long-dashed arrow does not appear to
describe {\it any} of the trends even remotely satisfactorily for ages
up to $t = 10^8$ yr. Specifically, we can rule out a 90 per cent
disruption rate per decade of age up to an age of 1 Gyr at the
$\gtrsim 8\sigma$ level (where $\sigma$ refers to the Poissonian
uncertainties shown in Fig. \ref{dndt.fig}).

As additional support of this conclusion, in Fig. \ref{dndt_comp.fig}
we reproduce the LMC cluster age distribution based on the full de
Grijs \& Anders (2006) sample (Fig. \ref{dndt.fig}), and add the
equivalent distributions based on both the Popescu et al. (2012) and
the Glatt et al. (2010) catalogues, using the same selection
limit. Although the Glatt et al. (2010) database includes stars down
to $V \sim 24$ mag, their LMC stellar census used to construct cluster
CMDs is significantly incomplete below $V \simeq 23$ mag (which limits
their cluster age determinations to a maximum of $\sim 1$ Gyr). This
implies that many low-mass, low-luminosity clusters are likely yet to
be detected. The census of brighter, more massive clusters is
significantly more complete and comparable among all studies
(cf. Baumgardt et al. 2013). However, since in this paper we will
apply the same selection limits to the Glatt et al. (2010) data as to
the de Grijs \& Anders (2006) and Popescu et al. (2012)
samples,\footnote{To convert the apparent magnitudes of Glatt et
  al. (2010) to absolute magnitudes, we adopted the canonical LMC
  distance modulus of $(m-M)_0 = 18.50$ mag.} the results should be
comparable for the appropriate age ranges. Recall that Glatt et
al. (2010) only considered clusters aged between $\sim 10$ Myr and 1
Gyr, but note as a caveat that for the youngest ages in the catalogue
the cluster census may be somewhat incomplete owing to a potentially
variable lower-age limit (see Section \ref{data.sec}). These authors
provide ages, extinction values and integrated $V$-band photometry for
all clusters in their sample; although they do not state specifically
whether their $V$-band magnitudes have been extinction-corrected, our
interpretation of their description is that they are (but this makes a
negligible difference to our results, in any case).

Reassuringly, all three distributions exhibit the same overall
behaviour and even their absolute scaling renders the distributions
virtually indistinguishable. All three samples, based on two
independent photometric catalogues and three independently determined
age distributions, are consistent with an age distribution for ages up
to $\sim 10^8$ yr that is best described by simple stellar evolution
(i.e., evolutionary fading), without the need for additional
disruption. In all three cases, the slope of the distribution becomes
significantly steeper only for ages in excess of 100 Myr, where we are
likely witnessing the onset of dynamical disruption (cf. Boutloukos \&
Lamers 2003; Lamers et al. 2005).

If we now compare Figs \ref{dndt.fig} and \ref{dndt_comp.fig} with
fig. 17 (left) of Chandar et al. (2010a), we first note that for ages
$\gtrsim 10^8$ yr, the overall distributions appear fairly similar,
with a significant steepening of the distribution occurring around 100
Myr. However, for younger ages, the Chandar et al. (2010a)
distribution is `negatively curved', compared with the `positively
curved' age distributions resulting from the three catalogues
considered in this paper. In fact, this appearance is predominantly
driven by Chandar et al.'s (2010a) youngest age bin, which exhibits a
clear excess in cluster numbers compared with the youngest age bin in
the other distributions discussed in this context. We will explore the
background to this apparent discrepancy in the next section.

\subsection{Discrepancies}
\label{discrepancies.sec}

\begin{figure}
\psfig{figure=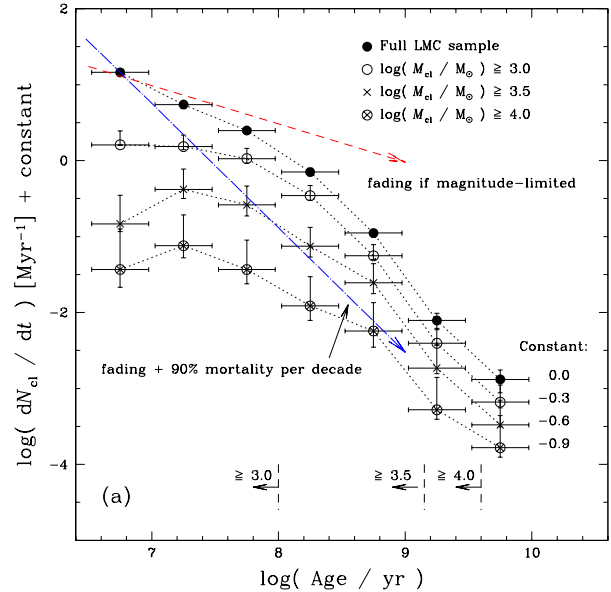,width=\columnwidth}
\caption{\label{dndt.fig}LMC cluster age distribution expressed as
  number of clusters per Myr. Shown are four different samples,
  including the full magnitude-limited LMC sample, and three
  mass-limited subsamples (shifted vertically, for reasons of clarity,
  by the constant offsets indicated). The mass-limited subsamples are
  50 per cent complete to the left of the vertical dashed lines
  included at the bottom of the figure, where the numbers refer to the
  50 per cent completeness limits for a given range, expressed in
  $\log(M_{\rm cl}/{\rm M}_\odot)$. The vertical error bars are
  Poissonian errors; the horizontal error bars indicate the age ranges
  used for the generation of these data points. The dashed arrow shows
  the expected effects due to evolutionary fading of a cluster sample
  made up of SSPs, based on the {\sc galev} SSP models, while the
  dash-dotted arrow represents the combined effects of a fading
  cluster population and 90 per cent cluster disruption per decade in
  $\log( t \mbox{ yr}^{-1})$.}
\end{figure}

\begin{figure}
\psfig{figure=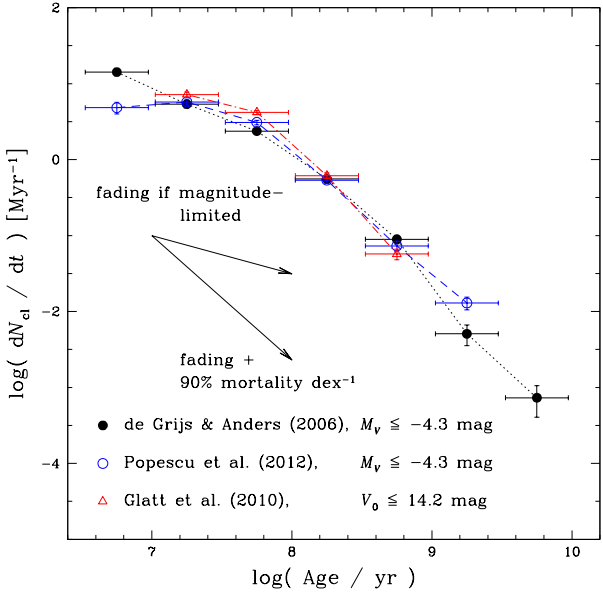,width=\columnwidth}
\caption{\label{dndt_comp.fig}As Fig. \ref{dndt.fig}, but for
  magnitude-limited subsamples based on all three catalogues discussed
  in this paper.}
\end{figure}

Note that one of the main differences between our results and those of
Chandar et al. (2010a,b) is driven by our respective analysis
methods. Chandar et al. (2010a,b) characterize their entire age- and
mass-limited cluster (sub-)samples by a single disruption law (i.e., a
straight-line fit in their equivalent representations of our
Fig. \ref{dndt.fig}), which may not be warranted, as we show here. Our
approach, on the other hand, is to explore whether a 90 per cent
disruption rate is supported for the earliest age ranges. The
differences between both sets of results therefore hinge on the
treatment of the data points pertaining to the youngest ages.

We will, therefore, perform a detailed comparison of Chandar et al's
(2010a) age--mass diagram (their fig. 3; top panel) with the
equivalent diagrams shown in Fig. \ref{amd.fig}. The main differences
in the cluster distributions between de Grijs \& Anders (2006) and
Chandar et al. (2010a) are (i) the presence of a population of young,
high-mass clusters in the Chandar et al. (2010a) data set, which are
virtually absent in the de Grijs \& Anders (2006) results, and (ii) an
overdensity (chimney) of clusters near $\log( t \mbox{ yr}^{-1} )
\simeq 6.6$ in the Chandar et al. (2010a) data. The latter overdensity
is related to the fitting procedure, as already acknowledged by
Chandar et al. (2010a) in the context of their comparison with the
original Hunter et al. (2003) results. Similar overdensities, although
for different ages, are seen in our age--mass diagram of
Fig. \ref{amd.fig}a (cf. Section \ref{data.sec}). This type of
behaviour is inherent to the use of broad-band SEDs to determine
integrated cluster properties.

The significant difference in the number of young, high-mass clusters
between both studies is more worrying: such objects are among the
brightest sources in a given cluster sample and should therefore be
found in any analysis. To explore the reason for this discrepancy, we
specifically focus on the section of parameter space covered by $\log(
t \mbox{ yr}^{-1}) \le 6.6$ and $\log( M_{\rm cl}/{\rm M}_\odot ) \ge
3.5$, indicated by the blue dashed boxes in Figs \ref{amd.fig}a and
\ref{amd.fig}b. Since the parameters derived by Chandar et al. (2010a)
are not publicly available, we base our comparison on their published
figure. In the relevant section of parameter space, Chandar et
al. (2010a) include 18 objects in their fig. 3 (top panel). The
equivalent region contains a single source in de Grijs \& Anders
(2006), whereas in Popescu et al. (2012) this region remains
unoccupied. We re-emphasize that all of these studies used the same
photometric database as input for their cluster age and mass
distributions.

Chandar et al. (2010a) based their broad-band SED fits on the Bruzual
\& Charlot (2003) SSPs for $Z = 0.008$, a Salpeter (1955)-type IMF and
Fitzpatrick's (1999) Galactic extinction law. Although de Grijs \&
Anders (2006) and Popescu et al. (2012) used different SSP models and
extinction laws, these choices are not expected to lead to
significantly different cluster age and mass estimates (cf. de Grijs
et al. 2005). The main difference between the approach taken by
Chandar et al. (2010a) on the one hand and de Grijs \& Anders (2006)
and Popescu et al. (2012) on the other resides in the choice of
stellar IMF. The latter studies used a Kroupa (2002)-type IMF, which
would yield lower cluster masses by a factor of $\sim 3.8$ (or $\sim
0.6$ dex) compared to the use of a Salpeter (1955) IMF. However,
Chandar et al. (2010a) argue that this difference is offset by the
need to apply aperture corrections to the original integrated cluster
photometry, thus eventually leading to similar masses. Finally, we
note that the youngest isochrone in the {\sc galev} SSP models is
characterized by an age of $\log( t \mbox{ yr}^{-1}) = 6.6$, whereas
the youngest object in Chandar et al. (2010a) in the parameter space
of interest has an age of $\log( t \mbox{ yr}^{-1}) \simeq 6.26$ (1.8
Myr); we considered this too young for a cluster to have emerged from
its natal molecular and dust cloud (cf. Section \ref{cmf.sec}).

With these differences in mind, we reverse engineered the photometric
measurements in the Johnson $V$ band (with and without extinction
corrections) that would be associated with the Chandar et al. (2010a)
age/mass combinations for the youngest, highest-mass clusters located
in the dashed regions in Fig. \ref{amd.fig}. On the basis of a
comparison of the $V$-band magnitudes thus derived with the original
integrated cluster photometry, we conclude that
\begin{enumerate}
\item the majority of the objects in this region with cluster
  parameters derived by Chandar et al. (2010a) are up to 3 mag
  brighter than any of the clusters in the original photometric
  database; and
\item a handful of the youngest, lowest-mass clusters (parameters as
  derived by Chandar et al. 2010a) may have counterparts in the de
  Grijs \& Anders (2006) and Popescu et al. (2012) data sets, although
  in these instances the latter authors obtained best-fitting
  parameters corresponding to older, more massive clusters. Such
  discrepancies can be traced back to the well-known
  age--extinction(--metallicity) degeneracy and are not a real reason
  for serious concerns in the context of this paper.
\end{enumerate}

Thus, while some of the lower-mass clusters of Chandar et al. (2010a)
in the section of parameter space of interest could have counterparts
in our own and the Popescu et al. (2012) databases, we are unable to
identify the highest-mass clusters in Chandar et al. (2010a) in the
original data set that forms the basis for all three analyses. Yet, it
is this subsample of clusters that drives the controversy and the
conclusion that significant cluster disruption may affect the LMC
cluster sample from the youngest ages onwards. Based on the comparison
performed here, we are forced to conclude that, in retrospect, this
claim appears to be unwarranted. In the following, we will attempt to
place this conclusion on a firmer quantitative footing.

\subsection{Cumulative distribution functions}

For any of the mass-limited subsamples, significant disruption does
not occur until $t \gtrsim 10^8$ yr (cf. Parmentier \& de Grijs 2008;
Baumgardt et al. 2013); for the full LMC cluster sample, one could
argue that some effects of disruption, in addition to evolutionary
fading, start to appear for $\log(t \mbox{ yr}^{-1}) \gtrsim
7.5$. Note that this age range is beyond that where we would consider
the relevant disruption process `infant' mortality (although this is a
matter of semantics).

Using a Monte Carlo approach, we will now attempt to quantify the rate
of disruption allowed by the data, using the cumulative cluster age
distribution. Our basic assumption is that clusters are born uniformly
in (linear) time following a power-law MF, $N(M_0) \propto
M_0^{-\alpha}$, where $\alpha = 2$. Irrespective of the effects of
infant mortality, if any, clusters evolve owing to stellar evolution
and two-body relaxation according to the formalism of Lamers et
al. (2005). The fraction of the mass of a cluster with initial mass
$M_0$ that is still bound at age $t$ is given by
\begin{equation}
\mu_{\rm ev}(t) \equiv \frac{M_{\rm cl}(t)}{M_0} = 1 - q_{\rm
  ev}(t),
\end{equation}
where $q_{\rm ev}$ is the mass lost through stellar evolution,
\begin{equation}
\log q_{\rm ev} = (\log t - a_{\rm ev})^{b_{\rm ev}} + c_{\rm ev},
\end{equation}
and $a_{\rm ev}=7, b_{\rm ev}=0.255$ and $c_{\rm ev}=-1.820$. The
`tidal' parameter $t_0$, defined as the normalization factor of the
disruption time-scale,
\begin{equation}
t_{\rm dis} = t_0 \Bigl( \frac{M_{\rm cl}}{M_\odot} \Bigr)^\gamma
\end{equation}
(and $\gamma = 0.67$ in our model; cf. Boutloukos \& Lamers 2003), can
be varied, but it only makes a measurable difference to our results if
$t_0 < 10^6$ yr, which has been ruled out by prior analysis of the LMC
cluster sample, which also established that $t_{\rm dis,LMC} > 10^9$
yr for $10^4$ M$_\odot$ clusters (Parmentier \& de Grijs
2008). Therefore, the cluster mass at age $t$ will simply be a
function of $M_0$ and age, modulo the $t_0$ parameter.

\begin{figure}
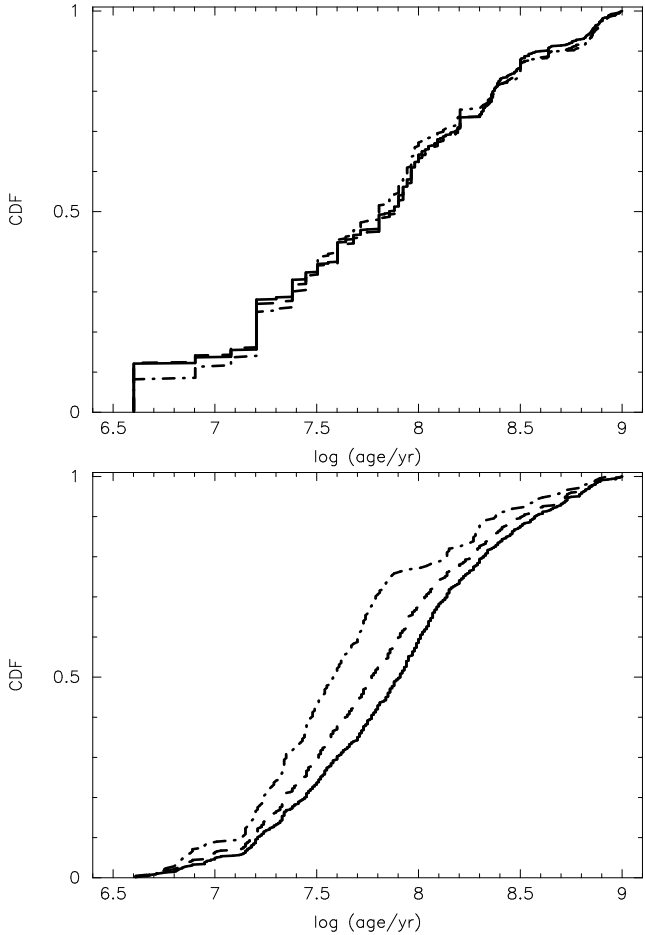

\psfig{figure=sg_fig2.ps,width=\columnwidth,angle=-90}
\psfig{figure=sg_fig3.ps,width=\columnwidth,angle=-90} 
\caption{\label{cumulobs.fig}Top: Cumulative distribution functions
  (CDFs) based on the de Grijs \& Anders (2006) cluster sample for
  `selection factors' $S = 0, 0.5$ and 1 -- as defined in
  Eq. (\ref{selfac.eq}) -- represented by the solid, dashed and
  dash-dotted lines, respectively. Bottom: As the top panel, but for
  the Popescu et al. (2012) sample.}
\end{figure}

\begin{figure}
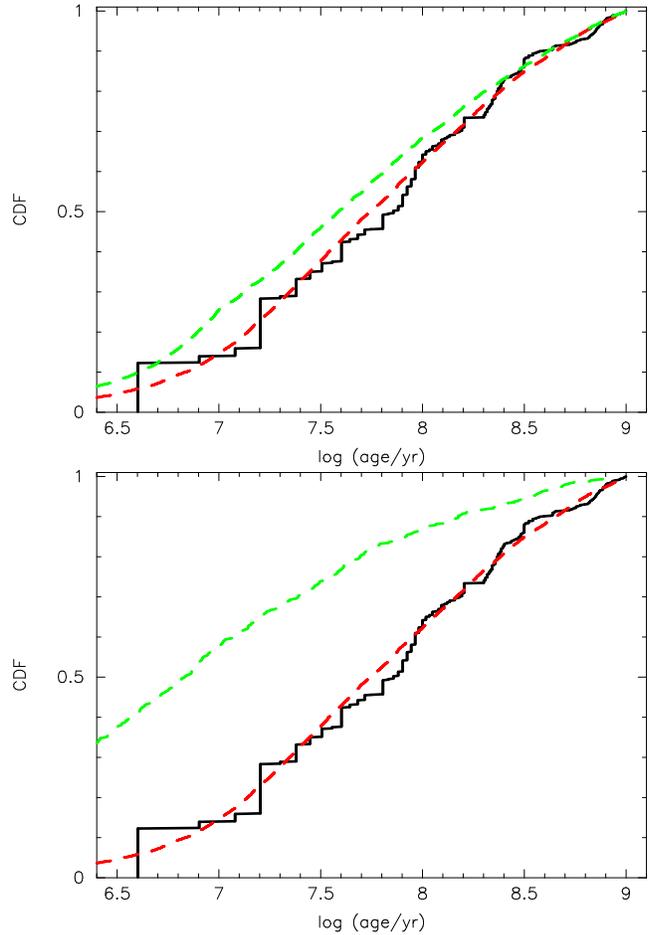

\psfig{figure=f4.ps,width=\columnwidth,angle=-90}
\psfig{figure=f5.ps,width=\columnwidth,angle=-90}
\caption{\label{cumul.fig}CDFs of the LMC cluster age
  distribution. The solid lines represent the de Grijs \& Anders
  (2006) data for $S = 0$; the red dashed lines are the best fit for
  $t > 4 \times 10^6$ yr (our lower boundary), assuming no disruption
  (specifically, no infant mortality), again for $S = 0$. Green dashed
  lines: Predicted CDFs for 50 per cent infant mortality at an age of
  10~Myr, followed by stellar evolutionary and dynamical evolution
  from the models of Lamers et al. (2005), adopting $t_0 = 10^6$~yr
  (top), and 90 per cent mortality per decade in $\log( t \mbox{
    yr}^{-1})$ and minimal evolution, i.e., $t_0 = 10^9$~yr (bottom),
  for $t \le 10^9$ yr, normalized at $t = 10^9$ yr.}
\end{figure}

Clusters are observed if their mass at a particular age is higher than
some limiting mass, $M_{\rm lim}(t)$ (defined by the observational
completeness limit), which is well represented by
\begin{equation}
\log(M_{\rm lim}/{\rm M}_\odot) = 0.5 \log(t \mbox{ yr}^{-1}) - 1.5 +
S,
\label{selfac.eq}
\end{equation}
where $S$ is a `selection factor' which allows us to explore the
importance of varying the selection limit: $S=0$ roughly follows the
selection limit of the observations for $6.5 < \log(t \mbox{ yr}^{-1})
< 9$ (see Fig. \ref{amd.fig}), while for $S=0.5$ this limit moves up
by $\Delta \log(M_{\rm cl}/{\rm M}_\odot) = 0.5$.

Figure \ref{cumulobs.fig} shows the LMC clusters' cumulative
distribution functions (CDFs) for (top) the de Grijs \& Anders (2006)
catalogue and (bottom) the Popescu et al. (2012) database using $S =
0, 0.5$ and 1 as selection limits. In the top panel, the CDFs contain,
respectively, 709, 544 and 256 clusters; the equivalent numbers in the
bottom panel are 671, 473, and 211 clusters,
respectively. Interestingly, the de Grijs \& Anders (2006) CDFs for
different values of $S$ are essentially the same (we will discuss the
differences seen in the Popescu et al. 2012 data below). There is no
apparent correlation between the numbers of clusters in any age range
with cluster mass (although the total numbers change), which is
consistent with a roughly constant cluster-formation rate. In
addition, if any significant level of cluster disruption were at play,
the data are also consistent with no strong mass dependence.

Figure \ref{cumul.fig} shows the CDF of the LMC cluster population
based on the de Grijs \& Anders (2006) sample (solid line, where the
sudden jumps are caused by the chimneys in the data set). The red
dashed lines represent the CDF (for $S = 0$) of the artificially
generated Monte Carlo clusters characterized by a constant
cluster-formation rate and based on standard $N$-body dynamics,
including the effects of stellar evolution but no infant
mortality. Note that in all cases where we show CDFs, here and below,
the data and model must necessarily match at a cumulative fraction of
unity, which represents our normalization. The observations are well
fitted by a model without the need for any significantly enhanced
(infant) mortality after 4 Myr, beyond the disruption that would be
expected from the combined action of stellar evolution and stellar
dynamics (predominantly two-body relaxation) on time-scales up to
$10^9$ yr. Note that we model such evolution using the analytic
prescription of Lamers et al. (2005). In reality, the $t_0$ parameter
would vary with position and time, which may lead to some (possibly
significant) differences to the cluster disruption properties and
time-scales at different galactocentric radii (see, e.g., Bastian et
al. 2012). A small amount of additional disruption (either infant
mortality or dynamical dissolution) could be accommodated by the data,
but there is no need to do so. Where our models include the effects of
infant mortality, this is implemented by the removal of a given
fraction (as specified in the text) of the star cluster population at
an age of 10 Myr, irrespective of cluster mass.

The top panel of Fig. \ref{cumul.fig} also shows the expected CDF for
50 per cent cluster infant mortality at an age of 10 Myr, followed by
standard stellar and dynamical evolution, for the same $t_0$. Although
$t_0$ is a free parameter, the observational data are inconsistent
with significant early disruption, irrespective of the value of $t_0$:
in essence, early (infant) mortality causes a change in the CDF slope
at early times, which remains. The smooth shape of the data suggests
no `kink' and, therefore, no significant infant mortality.

The bottom panel of Fig. \ref{cumul.fig} explores the idea of a 90 per
cent disruption rate per decade in $\log(t \mbox{ yr}^{-1})$ up to $t
= 10^9$ yr. The green dashed line is the closest that this model
(i.e., for $t_0 = 10^9$ yr) is found to approach the data, given that
the data and the model must reach a cumulative fraction of unity at
the same time; it is clearly a very poor match. The main problem with
this model is that to find any clusters at all at the oldest ages
requires very large numbers of clusters at young ages, as shown by the
difference between the green dashed line and the data at the youngest
ages: instead of the observed fraction of $<15$ per cent, more than
half of our sample clusters would need to be younger than 10 Myr. If
this were the case, this would suggest significant deviations from the
roughly constant cluster formation rate implied by the observational
data (e.g., Maschberger \& Kroupa 2011; see also Section
\ref{bigpicture.sec}).

In the context of our comparison with the Chandar et al. (2010a,b)
results, the de Grijs \& Anders (2006) database is the most
appropriate comparison sample, given that it is also based on SED fits
assuming fully sampled cluster stellar IMFs. This notion is supported
by Chandar et al.'s (2010a,b) conclusion that the impact of sample
incompleteness on their results is minimal at the low-mass end, as we
also found for the de Grijs \& Anders (2006) sample. The effects
caused by stochastic sampling of the stellar MFs become apparent in
the bottom panel of Fig. \ref{cumul.fig}. Whereas the Popescu et
al. (2012) data describe a qualitatively similar behaviour as those in
the top panel of that figure in terms of the absence of a clear need
for significant cluster disruption to have occurred in the last $\sim
100$ Myr, taking into account stochastic sampling tends to lead to an
overproduction of massive clusters aged between approximately 10 and
30 Myr, compared to the results from `standard' modelling. This effect
becomes apparent for $S = 1$, i.e., well above our selection limit;
for $S = 0$ and 0.5, the Popescu et al. (2012) results are, in fact,
similar to (although not the same as) the de Grijs \& Anders (2006)
CDFs.

\subsection{A variable cluster-formation rate?}

\begin{figure}
\psfig{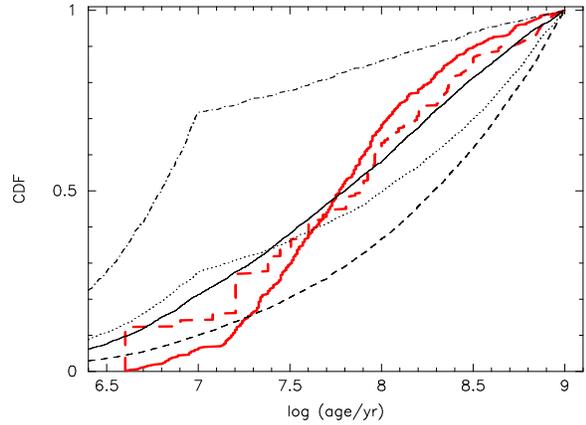}
\caption{\label{cumul-models.fig}CDFs of the LMC cluster age
  distributions based on both data sets considered in this paper (for
  $S=0$). The thick red lines represent the data sets (solid line:
  Popescu et al. 2012; dashed line: de Grijs \& Anders 2006); the
  black lines show various model predictions. All models adopt a
  power-law initial cluster MF with an index of $\alpha= 2$. From top
  to bottom, the models include (dash-dotted line) a scenario of 90
  per cent mass-independent infant mortality at 10 Myr, followed by
  stellar evolution and dynamical cluster disruption characterized by
  $t_0 = 10^6$ yr; (solid line) a scenario based on a constant
  cluster-formation rate and dynamical cluster disruption following
  Lamers et al. (2005) for $t_0 = 10^6$ yr; (dashed line) the same
  model as represented by the solid line, but for $t_0 = 10^7$ yr;
  (dotted line) a model showing the predictions for 70 per cent
  mass-independent infant mortality at 10 Myr, followed by stellar
  evolution and dynamical cluster disruption characterized by $t_0 =
  10^7$ yr.}
\end{figure}

\begin{figure}
\psfig{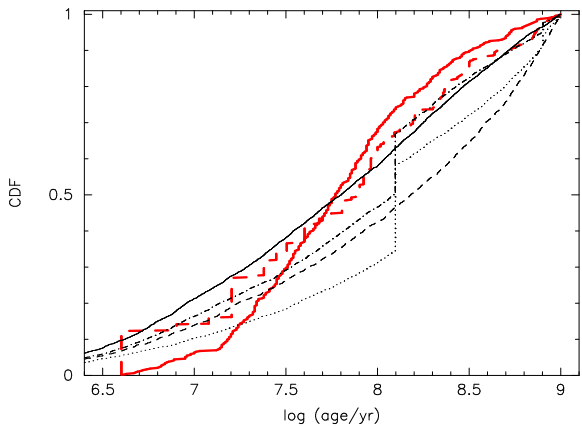}
\caption{\label{cumul-models2.fig}CDFs of the LMC cluster age
  distributions based on both data sets considered in this paper
  (again for $S=0$), exploring the effects of varying the
  cluster-formation rate. The thick red lines represent the data sets
  (solid line: Popescu et al. 2012; dashed line: de Grijs \& Anders
  2006); the black lines show various model predictions. All models
  adopt a power-law initial cluster MF with an index of $\alpha= 2$
  and dynamical cluster disruption following Lamers et al. (2005) for
  $t_0 = 10^6$ yr. From top to bottom, the models include (solid line)
  the reference scenario (identical to the solid line in
  Fig. \ref{cumul-models.fig}) based on a constant cluster-formation
  rate; (dash-dotted line) a constant cluster-formation rate in linear
  time with two additional bursts of cluster formation at 125 and
  800~Myr, each of which formed 10 per cent of the total number of
  clusters; (dashed line) a continuously decreasing cluster-formation
  rate (in linear time) by a factor of four between 1~Gyr and the
  present time; (dotted line) a decreasing cluster-formation rate (in
  linear time), but with the addition of two bursts at 125 and
  800~Myr. The latter model is hence a combination of the dash-dotted
  and dashed lines.}
\end{figure}

Figures \ref{cumul-models.fig} and \ref{cumul-models2.fig} illustrate
and summarize the expected effects of cluster infant mortality and of
changing the cluster-formation rate and the characteristic disruption
time-scale, $t_0$. Both figures show the de Grijs \& Anders (2006) and
Popescu et al. (2012) data sets for $S=0$ (thick dashed and solid red
lines, respectively), as well as four representative models each
(black lines of different styles). All models adopt a power-law
initial cluster MF with an index of $\alpha= 2$, and dynamical cluster
disruption following the Lamers et al. (2005) prescription. The models
in Fig. \ref{cumul-models.fig} are based on a constant
cluster-formation rate, combined with cluster disruption for $t_0 =
10^6$ yr and $t_0 = 10^7$ yr (black solid and dashed lines,
respectively). The black dash-dotted line represents a scenario of 90
per cent mass-independent infant mortality at 10 Myr, followed by
stellar evolution and dynamical cluster disruption characterized by
$t_0 = 10^6$ yr (cf. Fig. \ref{cumul.fig}. Similarly, the black dotted
line is for 70 per cent mass-independent infant mortality at 10 Myr
and disruption characterized by $t_0 = 10^7$ yr.

A straight line in this diagram would represent equal numbers of
clusters per decade in age. This is, in essence, shown by the black
solid line, combined with a short characteristic disruption time-scale
($t_0 = 10^6$ yr). The black dashed line in
Fig. \ref{cumul-models.fig} is relevant for a population containing a
larger number of older clusters (closer to equal numbers per linear
time period), since $t_0$ is longer, at $10^7$ yr. If the LMC cluster
population were characterized by a scenario in which 90 per cent of
clusters had suffered from infant mortality (as suggested by Chandar
et al. 2010a,b), the vast majority of clusters would need to be young,
since only 10 per cent would survive and contribute to the observed
CDF.

The alternative to the scenario represented by the black solid line
(no infant mortality and a short characteristic disruption
time-scale), would be evolutionary conditions dominated by $t_0
  \sim 10^7$ yr and at most a small amount of infant mortality
  acting on time-scales up to a few $\times 10^7$ yr. The dotted line
  in Fig. \ref{cumul-models.fig} is based on the assumption that all
  infant mortality has occurred by $t = 10^7$ yr. Infant mortality
  which acts by an age of a few $\times 10^7$ yr, as usually adopted,
  causes a clearly discernible kink in the CDF, which in turn causes
  the model to attain values that are significantly too large at the
  youngest ages and not seen in either of our data sets. As such,
  neither of our data sets support a significant amount of infant
  mortality at early times ($t \lesssim$ a few $\times 10^7$ yr).

Figure \ref{cumul-models2.fig} shows the effect on the CDF of changing
the cluster-formation rate. The thin black lines show four different
scenarios for the LMC's cluster-formation history over the past Gyr
for a characteristic dissolution time-scale $t_0 = 10^6$ yr. None of
the models in this figure include any infant mortality; they only
include evolutionary fading of their stellar populations and
evaporation owing to dynamical evolution.

Increasing the past cluster-formation rate (through adoption of either
a generally higher rate or in a series of bursts) leads,
unsurprisingly, to an increase in the number of clusters at older
ages.  The effect of introducing bursts of cluster formation are more
or less obvious, depending on the fraction of the total number of
clusters formed in these bursts. As an example, we adopted a scenario
in which 10 per cent of the total number of clusters were formed in
each instantaneous burst. Although this is admittedly excessive, it
helps to illustrate the point we want to convey based on
Fig. \ref{cumul-models2.fig}.

It might na\"\i vely be thought that increasing the past
cluster-formation rate -- and so increasing the number of older
clusters relative to younger clusters -- would give scope for an
enhanced rate of infant mortality. However, in
Fig. \ref{cumul-models2.fig} the model CDFs for higher past
cluster-formation rates lie below the observations, whilst in
Fig. \ref{cumul-models.fig} including infant mortality, combined with
a constant cluster-formation rate, means that there are too few old
clusters compared to the observations. Let us now consider what this
means for a scenario of `classical' infant mortality. If infant
mortality is rapid (i.e., caused by gas expulsion shortly after
cluster formation), then the infant-mortality-induced loss of clusters
from one's sample occurs after some 10--20 Myr. Therefore, it is at
the very youngest ages that clusters must be vastly overproduced
relative to the observations. Including a decreasing cluster-formation
rate with time lowers the (dash-dotted) 90 per cent infant mortality
model line in Fig. \ref{cumul-models.fig} to some extent, but still
over half of the clusters in our samples should be $<10$~Myr old
compared to the 15 per cent that is observed.

In summary, the relatively small number of young clusters in both the
de Grijs \& Anders (2006) and the Popescu et al. (2012) samples
implies that a scenario in which 90 per cent of the cluster population
undergoes infant mortality is unrealistic and not supported by either
data set. Alternatively, a longer characteristic disruption time-scale
appears to be ruled out as well, given that there are too few old
clusters in either of our catalogues to support such a model. In
addition, adoption of a longer disruption time-scale will cause the
CDF to be increasingly shifted to older ages, hence leading to ever
more significant model {\it under}predictions compared with the
behaviour of the actual data sets.

Finally, although we set out to show that a scenario involving 90 per
cent cluster infant mortality at early times appears to be ruled out
by both data sets, we will now comment briefly on the shapes of the
CDFs defined by our two data sets. Although they are largely
consistent with one another for $S=0$, the data sets exhibit some
small systematic differences, in particular for the youngest clusters
($t \lesssim$ a few $\times 10^7$ yr). In the context of the
diagnostic age--mass diagram, we attributed this to the effects of
stochastic sampling of the clusters' stellar MFs. In relation to the
CDFs discussed in this section, these differences may be either
realistic or caused by a preferential reduction of cluster masses
based on stochastic modelling. We are, indeed, concerned that such a
bias may have been introduced by the stochastic modelling approach. We
are currently exploring these issues using our newly developed,
extensive stochastic model set based on the {\sc galev} stellar
population models (Anders et al. 2013). We will apply these models to
our unprecedented {\sl Hubble Space Telescope}-based imaging data set
of the rich star cluster system associated with the dwarf starburst
galaxy NGC 5253 (de Grijs et al. 2013), which covers 10 passbands from
near-ultraviolet to near-infrared wavelengths.

If, on the other hand, these systematic differences reflect the true
physical properties of the LMC cluster sample as derived by Popescu et
al. (2012), what would this mean for our analysis in this section?
The de Grijs \& Anders (2006) data set appears to be well represented
by a roughly constant cluster-formation rate over the time span
considered here; to match the Popescu et al. (2012) data set, a
scenario involving a slight enhancement or a minor burst in the
cluster-formation rate in the past few $\times 10^7$ yr might provide
a somewhat better match. One can, of course, vary past
cluster-formation rates and different prescriptions of cluster
mortality (infant or otherwise) to find a good fit to the
observations. For example, a generally decreasing cluster-formation
rate but with periods of enhanced cluster formation in the past
100~Myr provides a good fit without the need to invoke infant
mortality. Vastly increased cluster formation 50--300~Myr ago with
significant infant mortality can also produce a reasonable fit (even
though such a scenario predicts almost no 20--50~Myr-old clusters). An
acceptable fit can also be obtaind with significant cluster mortality
occurring at 100~Myr rather than 10~Myr. However, without a good
physical reason to think that these are reasonable models,
particularly in the absence of supporting evidence for such scenarios
based on independent studies (see also Section \ref{bigpicture.sec}),
we argue that it is rather pointless to pursue such fits.

\section{Context}

We have thus far specifically focussed on a detailed and thorough
(re-)analysis of the LMC cluster population as covered by the Hunter
et al. (2003) database. Since a number of different authors reached
conflicting conclusions as regards the early evolution of the galaxy's
cluster system, but based on the same basic photometric data set, our
aim was to explore the underlying reasons for this discrepancy. In
this section, we take the discussion further by addressing the more
general context associated with this work. In particular, we will
address (i) the key assumption that the LMC's cluster-formation
history has remained roughly constant over the past $\sim 1$ Gyr, and
(ii) the impact of the partial coverage of the LMC's extent by the
Hunter et al. (2003) cluster database on our understanding of the LMC
cluster population's properties and evolutionary history as a whole.

\subsection{The LMC's cluster-formation rate}
\label{bigpicture.sec}

Under the key assumption that the cluster formation rate has remained
roughly constant, our `null hypothesis', we concluded that there is no
compelling evidence (within the uncertainties) of significant cluster
disruption for $t \lesssim 100$ Myr. Our analysis of the shape and
normalization of the cluster MFs, aided by our results from an
assessment of the CDFs for different selection limits, supports the
notion of a roughly constant cluster-formation rate. In this section,
we will address the validity of this assumption, so as to place our
results in the more general context of the LMC's overall star- and
cluster-formation history.

Prior to the series of papers based on the Hunter et al. (2003)
photometric cluster database, the only `modern' analyses of the LMC's
cluster-formation history were published by Girardi et al. (1995) and
Pietrzy\'nski \& Udalski (2000). Girardi et al. (1995) analysed
integrated $UBV$ photometry from a pre-publication release of Bica et
al.'s (1996) catalogue and concluded that the LMC's evolutionary
history is characterized by periods of enhanced cluster formation, by
a factor of $\lesssim 2$, at $\sim 100$ Myr and 1--2 Gyr, as well as
by the well-known, pronounced `age gap' between $\sim 3$ and [12--15]
Gyr (see below). Pietrzy\'nski \& Udalski (2000) used ischrone fits to
determine ages of up to 1.2 Gyr of 600 clusters in the central LMC
(bar) area. They concluded that the LMC cluster-formation rate is
characterized by a number of bursts with complex age structure,
specifically centred at ages of $\sim 7$, 125 and 800 Myr. The most
recent period of enhanced cluster formation produced of order a factor
of 1.5--2 more clusters per unit (linear) age range than the
equivalent rate during the galaxy's more quiescent period(s), while
the burst centred at $\sim 125$ Myr produced cluster numbers boosted
by yet another factor of $\sim 2$ -- when we smooth the Pietrzy\'nski
\& Udalski (2000) cluster age distribution to the same resolution as
adopted in this paper.

However, the empirically derived cluster age distribution is the
product of cluster formation {\it and} disruption as a function of
time. Adopting the cluster age distribution as proxy of a galaxy's
cluster-{\it formation} history only yields, therefore, merely part of
the story. A roughly constant age distribution could therefore imply a
similarly shaped cluster-formation history, but it could also result
from a balanced interplay between cluster formation and
disruption.

A number of authors have suggested that the LMC's resolved stellar
population could provide clues as to the galaxy's cluster-formation
history based on CMD analysis (for a recent discussion, see
Maschberger \& Kroupa 2011), provided that the star- and
cluster-formation histories can be mapped onto one another within
reasonably small uncertainties. However, (massive) cluster and
field-star formation may well require different conditions to thrive
in, implying that the two formation scenarios may not always be
coincident. This type of scenario is likely, in fact, given the
observed disparities between the cluster and field-star age
distributions in, e.g., the Magellanic Clouds as well as in NGC 1569
(e.g., Anders et al. 2004a). In particular, the LMC exhibits a
well-known gap in the cluster age distribution, although the age
distribution of the field stellar population appears more continuous
(e.g., Olszewski et al. 1996; Geha et al. 1998; Sarajedini 1998; and
references therein). In addition, the cluster and field-star age
distributions are also significantly different in the SMC
(cf. Rafelski \& Zaritsky 2005; Gieles et al. 2007).

Maschberger \& Kroupa (2011) recently performed a very careful and
detailed study of the LMC's cluster-formation history (based on the
ages and masses from de Grijs \& Anders 2006) and its relationship, if
any, to the galaxy's field-star formation history. They compared the
LMC's star-formation history based on CMD analysis (using
observational data from Harris \& Zaritsky 2009) with the galaxy's
cluster-formation history resulting from consideration of both the
most massive clusters only (cf. Maschberger \& Kroupa 2007) and of the
total mass in clusters of any mass, although in the latter case they
could trace only the most recent [20--400] Myr period.

These authors found that the shape of the resulting cluster-formation
history matches that of the field-star formation history based on CMD
analysis very well for the past $10^9$ yr (cf. their fig. 8). The
absolute value for the star-formation rate based on their most-massive
cluster analysis also matches that of the field stars, while the
absolute values differ systematically for the results based on the
total mass in star clusters; this is interpreted in terms of either a
low bound star cluster-formation efficiency or a high degree of infant
mortality. Once again, therefore, these results based on empirical
cluster age distributions are affected by a degeneracy between cluster
formation and disruption scenarios.

The precise shape of the most recent LMC cluster-formation history
derived by Maschberger \& Kroupa (2011) depends ultimately on whether
or not the actively star-forming region centred on 30 Doradus (30 Dor)
and its massive central cluster R136 are included in the models. The
Hunter et al. (2003) catalogue does not contain the 30 Dor region (see
also the discussion in Section \ref{coverage.sec}). Maschberger \&
Kroupa (2011; their figs 3, bottom, and 4) show that the LMC's
cluster-formation rate over the past $\sim 1$ Gyr has remained
constant within $\sim 10$ per cent if 30 Dor is included, while it
shows a reduction in the cluster formation rate in the past few
$\times 10^7$ yr by a factor of 3--4 if 30 Dor is not included. The
cluster-formation history based on the total mass in clusters shows a
relatively enhanced period of cluster formation 20--40 Myr ago, and a
reduction by a factor of $\sim 3$--4 more recently (cf. their figs 5
and 6). Fig. 4 of Baumgardt et al. (2013), which shows the LMC cluster
system in the (${\rm d}N_{\rm cl}/{\rm d}t$ versus $\log t$) plane,
also supports this conclusion.

In summary, although the LMC's cluster-formation history is still
subject to sizeable uncertainties, our null hypothesis of a roughly
constant cluster-formation rate for the past $10^9$ yr is likely not
too far off the mark. There is little, if any, modern empirical
support for a significantly enhanced cluster-formation rate in the
past few $\times 10^7$ yr, by an order of magnitude or more, which
would be required to reconcile a high infant-mortality rate with the
results from our diagnostic tests in this paper, modulo the degeneracy
between cluster formation and disruption scenarios pointed out
above. If anything, the galaxy's cluster-formation rate may have
declined by a factor of a few compared with that 20--40 Myr ago.

\subsection{How representative is our LMC cluster data set?}
\label{coverage.sec}

Our results and the comparisons discussed in Sections 1 through 5 were
largely based on the Hunter et al. (2003) cluster photometry of 748
distinct star clusters above a nominal selection limit of $M_V \simeq
-4.3$ mag. To place these results into a more general context, we need
to consider whether and to what extent this cluster database is
representative of the LMC's cluster population as a whole.

The Hunter et al. (2003) cluster sample is based on Massey's (2002)
14.5 deg$^2$ CCD survey of the Magellanic Clouds; for an overview of
the survey's spatial coverage, see his fig. 1 (see also Maschberger \&
Kroupa 2011, their fig. 1). The Hunter et al. (2003) cluster sample
does not cover the entire LMC (e.g., it does not cover the entire bar
region), although this does not stop Hunter et al. (2003) from
specifically assuming that their objects are representative of the LMC
cluster population as a whole. Their database covers approximately
half of the LMC bar and a number of fields in the more extended LMC
disc region. One important caveat is that the actively star-forming
field centred on the 30 Dor region is not included in the
catalogue. As we saw in Section \ref{bigpicture.sec}, whether or not
the young, massive clusters in this field are included in our analysis
may lead to different interpretations as regards the shape of the
cluster-formation history.

Baumgardt et al. (2013, their fig. 1) show the coverage of the Hunter
et al. (2003) clusters with respect to that of both the Pietrzy\'nski
\& Udalski (2000; Optical Gravitational Lensing Experiment, {\sc ogle
  ii}) and the Glatt et al. (2010) samples. The {\sc ogle ii} sample
predominantly covers the central regions of the galaxy, including the
entire LMC bar. This sample thus extends the Hunter et al. (2003)
coverage to include the missing part of the bar region, but it does
not include a significant number of clusters in the more general field
of the LMC. The Glatt et al. (2010) data are based on the MCPS, which
covers the central 64 deg$^2$ of the LMC. Their clusters extend well
beyond the coverage of the Hunter et al. (2003) database, with a
particularly large excess of clusters towards the north compared to
the Hunter et al. (2003) coverage. Glatt et al. (2010) found evidence
of variable star (cluster)-formation histories across the LMC system,
so that a complete picture of the galaxy's cluster properties requires
the largest possible spatial coverage.

In the context of our focus on the youngest clusters in the LMC, it is
therefore particularly frustrating that the Glatt et al. (2010) sample
cannot shed light on the cluster formation and disruption scenarios
for clusters younger than a few $\times 10^7$ yr (cf. the discussion
in Section 2). Nevertheless, these authors state specifically that
``the youngest clusters reside in the supergiant shells, giant shells,
the intershell regions and towards regions with a high H$\alpha$
content''. These regions, in particular the (super-)giant shells as
well as the young, star-forming blue and south-eastern arms, are
mostly located outside the Massey (2002) survey area. However, in
Fig. \ref{dndt_comp.fig} we compared the Glatt et al. (2010) cluster
parameters with those of de Grijs \& Anders (2006) and Popescu et
al. (2012), {\it for the same limiting magnitude}, and concluded that
the ${\rm d}N_{\rm cl}/{\rm d}t$ distributions as a function of age
are both qualitatively and quantitatively similar for all
samples. Since adoption of a limiting magnitude is equivalent to
imposing a limiting mass for a given age, this leads us to suggest
that the Hunter et al. (2003) clusters -- which, after all, formed the
basis for the de Grijs \& Anders (2006) and Popescu et al. (2012)
results -- are representative of the LMC's cluster population at large
{\it for an age-dependent minimum mass limit} corresponding to $M_V
\simeq -4.3$ mag. In other words, many of the young clusters in the
Glatt et al. (2010) catalogue would fall below our selection limit
adopted here (see below). These results also suggest that, although
the Massey (2002) survey did not cover the entire LMC disc region, its
coverage is sufficient to trace a representative, magnitude
(mass)-limited sample of LMC clusters with ages up to 1 Gyr.

\begin{figure}
\psfig{figure=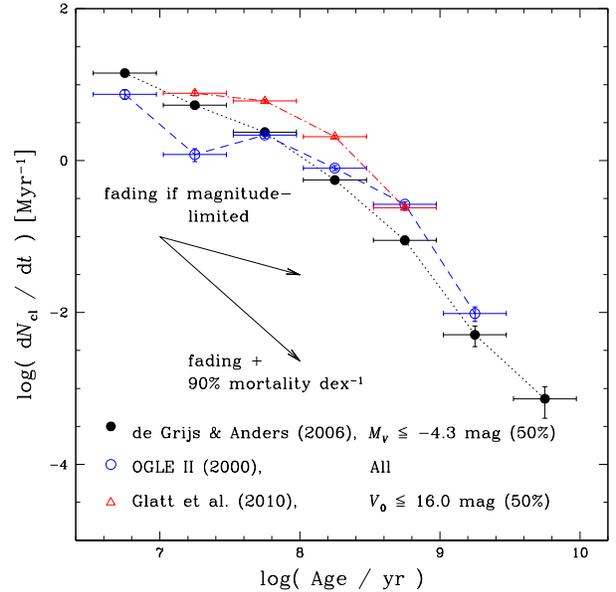,width=\columnwidth}
\caption{\label{dndt_lit.fig}As Fig. \ref{dndt_comp.fig}, but for the
  {\sc ogle ii} and Glatt et al. (2010) catalogues. The distributions
  include all clusters in the Pietrzy\'nski \& Udalski (2000) database
  of cluster ages based on {\sc ogle ii} and objects brighter than
  $V_0 = 16.0$ mag from the Glatt et al. (2010) sample, corresponding
  to the approximate 50 per cent completeness limit. The de Grijs \&
  Anders (2006) results, for $M_V \le -4.3$ mag, are shown for
  reference.}
\end{figure}

The diversity of cluster-formation rates across the LMC (cf. Glatt et
al. 2010) is exemplified by the ${\rm d}N_{\rm cl}/{\rm d}t$
distributions in Fig. \ref{dndt_lit.fig}. We show the de Grijs \&
Anders (2006) results for $M_V \le -4.3$ mag for reference. We also
include the Glatt et al. (2010) clusters, having imposed a limiting
magnitude of $V_0 = 16.0$ mag, which corresponds to their cluster
sample's approximate 50 per cent completeness limit, if we assume that
the bright end of the cluster luminosity function is adequately
represented by a single power law. This sample of clusters has been
drawn from across the extended LMC system, containing numerous
clusters outside the Massey (2002) fields (cf. Baumgardt et al. 2013;
their fig. 1). In addition, we show the distribution for the {\sc ogle
  ii} clusters centred on the LMC bar with isochrone-based age
determinations from Pietrzy\'nski \& Udalski (2000). The latter sample
is limited to clusters younger than about 1.2 Gyr because of the
observational completeness limit for single stars, $V \approx 21.5$
mag. The catalogue's photometric completeness characteristics have not
been explored in detail (cf. Pietrzy\'nski et al. 1999), but the depth
of the observations is of order 1.5 mag shallower than that of the
MCPS used by Glatt et al. (2010).

The differences in the cluster-formation (and, possibly, disruption)
histories among the three samples are clear. The Glatt et al. (2010)
extended LMC disc sample contains a significantly larger sample of
clusters at any age than the reference sample, but particularly for
$\log(t \mbox{ yr}^{-1}) \gtrsim 7.5$. Note, however, that the curve
turns down to lower rates more rapidly for younger ages than that
representing the de Grijs \& Anders (2006) sample. The {\sc ogle ii}
sample shows larger variations from one time step to the next, which
may imply a significantly more bursty cluster-formation rate in the
galaxy's bar (cf. Pietrzy\'nski \& Udalski 2000) than in the less
dense regions at larger radii sampled by both comparison samples. The
overall trend, however, does not support a significant increase of
cluster formation at very young ages.

Finally, we return to the obvious omission of the 30 Doradus region
and its massive, central star cluster R136. This is one of a very
small number of massive clusters left out of our analysis and which
would have been taken into account given the observational selection
limit imposed if it had been covered by the original survey
data. Addition of a single or a few young, $< 10^7$ yr-old clusters to
either of our main databases would increase the relevant number of
clusters in this age range by $\sim 10$--20 per cent. However, we
would need at least an order of magnitude more young clusters and a
factor of 2--4 more clusters with ages of $\log(t \mbox{ yr}^{-1})
\sim 7.5$ (and more massive than our age-dependent mass limit) to
conclusively support a high degree of early cluster disruption. The
empirical data do not allow us to reach such a conclusion.

\section{Conclusion}

On their own, the results based on any of the individual approaches
presented here are merely indicative of the physical conditions
governing the LMC's cluster population. However, the combination of
our results from all three different diagnostics leaves little room
for any conclusion other than that a high rate of early cluster
disruption is summarily ruled out. The CDF results show, in
particular, that high levels of infant mortality require that the vast
majority of one's cluster sample must be young, unless the
cluster-formation rate were significantly higher: for a disruption
rate of 90 per cent per decade in age up to $10^9$ yr, instead of the
observed fraction of $<15$ per cent, more than half of our sample
clusters would need to be younger than 10 Myr. Such high star- and
cluster-formation rates appear to be ruled out for $t \lesssim 10^9$
yr on the basis of analyses of both the cluster population
(Maschberger \& Kroupa 2011) and the LMC field's star-formation
history (e.g., Harris \& Zaritsky 2009; Rubele et al. 2012; and
references therein).

We thus conclude that the cluster disruption rate in the LMC, at least
over the past 100 Myr, has been well below that found for large and/or
interacting galaxies like the Antennae system (Whitmore, Chandar \&
Fall 2007; and references therein), M51 (Bastian et al. 2005), and the
Milky Way (Lada \& Lada 2003). We do not find any compelling evidence
of significant cluster disruption and estimate a conservative maximum
disruption rate of less than 10 per cent per decade in $\log( t \mbox{
  yr}^{-1})$, up to $t \simeq 10^8$ yr. It seems, therefore, that the
difference in environmental conditions in the Magellanic Clouds on the
one hand and significantly more massive galaxies on the other may be
the key to understanding the apparent variations in cluster disruption
behaviour.

\section*{Acknowledgements}
We thank Holger Baumgardt and the referee for helpful contributions
and suggestions, respectively. RdG and PA acknowledge research support
through grant 11073001 from the National Natural Science Foundation of
China (NSFC). This research has made use of NASA's Astrophysics Data
System Abstract Service.

\end{document}